\documentclass[12pt]{iopart}

\usepackage{epsfig}
\usepackage{amssymb}

\begin{document}

\letter{Non-equilibrium Lorentz gas on a curved space}
\author{Felipe Barra}
\address{Deptartamento de F\'{\i}sica, Facultad de Ciencias
  F\'{\i}sicas y Matem\'aticas, Universidad de Chile, Casilla 487-3,
  Santiago Chile}
\author{Thomas Gilbert}
\address{Center for Nonlinear Phenomena and Complex Systems,
  Universit\'e Libre  de Bruxelles, CP~231, Campus Plaine, B-1050
  Brussels, Belgium}
\date{\today}

\begin{abstract}
The periodic Lorentz gas with external field and iso-kinetic thermostat is
equivalent, by conformal transformation, to a billiard with expanding
phase-space and slightly distorted scatterers, for which the trajectories
are straight lines. A further time rescaling allows to keep the speed
constant in that new geometry. In the hyperbolic regime, the stationary
state of this billiard is characterized by a phase-space contraction rate,
equal to that of the iso-kinetic Lorentz gas. In contrast to the
iso-kinetic Lorentz gas where phase-space contraction occurs in the bulk,
the phase-space contraction rate here takes place at the periodic
boundaries.  
\end{abstract}

\submitto{J. Stat. Mech.}


\ead{fbarra@dfi.uchile.cl, thomas.gilbert@ulb.ac.be}

\maketitle

\nosections

Over the last twenty years, the study of time-reversible dissipative
systems with chaotic dynamics as mechanical models of non-equilibrium
stationary processes has played an important role in shaping our
understanding of the connections between irreversible macroscopic processes
and the reversible dynamics that underly them at the microscopic level, see
\cite[and refs. therein]{D99}.  Among other large deviation
relations, the Fluctuation Theorem \cite{ECM93,GC95} is a central result of
this approach. Related results have appeared since then \cite{J97,C99},
which have been put to the test experimentally in small circuits and
biological molecules, see \cite[and refs. therein]{BLR05}. 

The forced periodic Lorentz gas with Gaussian iso-kinetic thermostatting
was first proposed by Moran and Hoover \cite{MH87} as such a
time-reversible dissipative chaotic model for conduction, where
irreversibility manifests itself in the fractality of phase-space
distributions. 

A rigorous analysis of that system was later provided by Chernov {\em et
  al.} \cite{CELS93}, establishing a relation between the sum of the
Lyapunov exponents and the entropy production rate of this system. The
thermostat is here a mechanical constraint, chosen according to Gauss'
principle of least constraint, which acts so as to remove the energy input
from an external field \cite{EM90}. Under this constraint, the kinetic
energy  remains constant and thus fixes the temperature of the system; no
interaction with a hypothetical environment is needed in order to achieve
thermalization. Rather the Gaussian thermostat causes dissipation in the
bulk. Due to the presence of a regular array of circular scatterers on
which non-interacting particles collide elastically, this system sustains a
chaotic regime, at least so long as the external forcing is not too
strong. As shown in \cite{CELS93}, this model has a unique natural
invariant measure, with one positive and one negative Lyapunov exponents,
whose sum is negative --a signature of the fractality of the invariant
measure-- and identified as minus the entropy production rate 
\cite{R96}. The comparison with  the corresponding phenomenological
expression provides a relation between the phase-space contraction rate and
conductivity.  

As will be reviewed shortly, the trajectories of the iso-kinetic billiard
are integrable from one collision to the next. It was shown by Wojtkowski
\cite{W00} that there exists a conformal transformation on the Complex
plane that transforms those curved trajectories into straight
lines. Formally, this amounts to introducing a metric, specified according
to the conformal transformation, which yields an identification between the
Gaussian iso-kinetic trajectories and the geodesics of a torsion free
connection, called the Weyl connection, for which the metric is preserved
under parallel transport. This is a natural 
generalization of the symplectic formalism for the Gaussian iso-kinetic
motion introduced by Dettmann and Morriss \cite{DM98}. The corresponding
billiard, whose trajectories follow the geodesics of the Weyl connection,
and undergo elastic collisions when they reach the boundary,
is referred to as the billiard W-flow. For our purpose, it will be
sufficient to focus our attention on the conformal transformation itself,
leaving aside the formal aspects of this geometric construction, which we
will henceforth loosely refer to as Weyl geometry.

In what follows, we will present the details of this conformal
transformation and study the dynamics of the billiard. Our main observation
is that the bulk dissipation of the iso-kinetic Lorentz billiard, which
accounts for the positive entropy production rate, disappears under the
conformal map, after time reparametrization. The phase-space contraction
rate and  its identification with the entropy production rate are
nevertheless recovered because of the periodic boundary conditions, which,
under the conformal map, induce a phase-space contraction rate which, in
average, is equal to the bulk dissipation of the iso-kinetic Lorentz
billiard. We mention that a billiard with a similar mechanism of
contraction of phase-space volumes at the borders was considered in
\cite{BR01}. However the connection to iso-kinetic dynamics was not
discussed there.

\begin{figure*}[htb]
\begin{center}
\includegraphics[width=.6\textwidth]{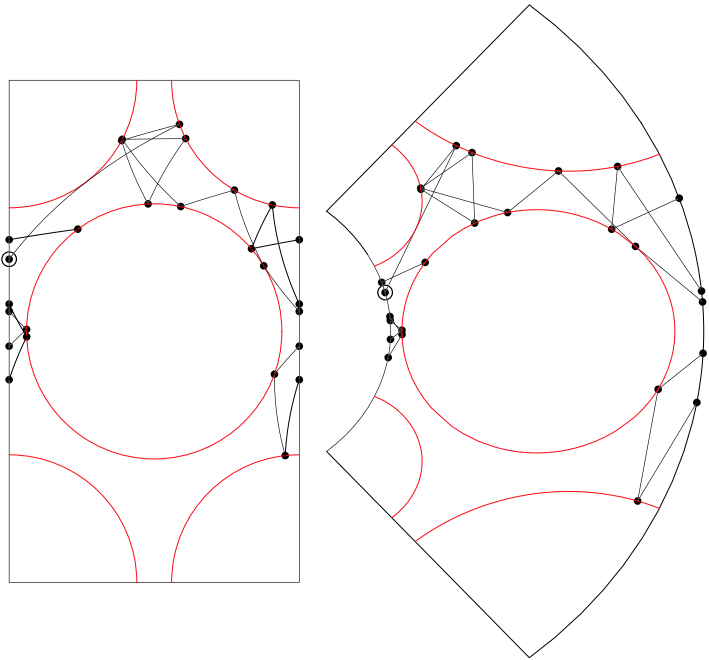}
\end{center}
\caption{(left) Unit cell with a typical trajectory, corresponding to
  the external field $\epsilon=1$; (right) The image of that cell and
  trajectory under the conformal map (\ref{confmap}). The initial condition
  is marked by a circled dot.}
\label{fig.gik}
\end{figure*}

We consider the two-dimensional iso-kinetic periodic Lorentz channel with
constant external field of magnitude $\epsilon$. Due to the periodicity of
this system, the dynamics can be studied in a unit cell with periodic
boundary conditions, as displayed on the left panel of
Fig. \ref{fig.gik}. This cell is an open rectangular domain, centered at 
the origin, with one disk at the cell's center, and four others at its
summits $(\pm 1/2, \pm\sqrt{3}/2)$. 
All the disks have identical radii, which we denote by $\sigma$, chosen so
as to satisfy the finite horizon condition, {\em i.~e.}
$\sqrt{3}/4 < \sigma<1/2$. The width of the cell is here taken to be unity,
and its height $\sqrt{3}$. Periodic boundary conditions apply at $x=\pm1/2$
and reflection at $y=\pm\sqrt{3}/2$. The direction of the external field is
taken towards the positive $x$ direction. Thus particles typically tend to
move in that direction, winding around the cell from one boundary
to the other. 

Let $(x_t,y_t)$ denote the coordinates of the particle at time $t$, and let
$\theta_t$ denote the 
velocity angle, measured with respect to the $x$-axis. Assume $p^2 \equiv
{\dot x_t}^2 + {\dot y_t}^2 \equiv 1$, which we take as the definition of
the temperature. In between collisions the equations of motion are
specified by
\begin{eqnarray}
\dot x_t &=& \cos\theta_t,\label{eqx}\\
\dot y_t &=& \sin \theta_t,\label{eqy}\\
\dot \theta_t &=& - \epsilon \sin \theta_t,\label{eqtheta}
\end{eqnarray}
with the solutions 
\begin{eqnarray}
x_t &=& x_0 + \frac{1}{\epsilon}\log\frac{\sin\theta_0}{\sin\theta_t},
\label{x}\\
y_t &=& y_0 + \frac{\theta_0-\theta_t}{\epsilon},\label{y}\\
\theta_t &=& 2\arctan\left[ \tan
  \frac{\theta_0}{2}\exp(-\epsilon t)\right]. 
\label{theta}
\end{eqnarray}
When the trajectory collides with a disk, $\theta_t$ changes according to the
rules of elastic collisions.

Let $z = x + \imath y$ and  $f(z)= -\epsilon (x+\imath y)$. That is, $f$ is
a holomorphic function whose real part is specified by the potential
associated to the external field. According to Wojtkowski \cite{W00}, the
conformal mapping 
\begin{equation}
F(z) = \int \exp[-f(z)]dz,
\label{confmap}
\end{equation}
takes the trajectories (\ref{x}, \ref{y}) into straight lines. This is
to say $F(t) \equiv F(z_t)$ is a straight line in the complex plane. 
Furthermore, since the transformation is conformal, the elastic collisions
are mapped into elastic collisions. The periodic boundary conditions are
modified under this map, as will be discussed shortly.  

The map (\ref{confmap}) takes a point with coordinates $(x,y)$ to the point
$(u,v)$ in the complex plane, 
\begin{eqnarray}
x + \imath y \mapsto
u + \imath v &=& F(x+\imath y),\nonumber\\
 &=& \frac{1}{\epsilon} 
\{\exp[\epsilon(x + \imath y)] - 1\}.
\label{confmapexp}
\end{eqnarray}
Note that the term $-1/\epsilon$ was introduced so as to keep the origin 
fixed under the transformation.

We show that trajectories are straight lines in the $(u,v)$ plane and
identify their slopes. To that end, we compute the time derivative of
$F(z_t)$ with  $z_t=x_t+\imath y_t$. From Eqs.~(\ref{eqx}-\ref{eqtheta}),
\begin{eqnarray}
\frac{d}{dt}F(z_t) &=& \exp[ \epsilon (x_t + \imath y_t)] 
\frac{d}{dt}(x_t+\imath y_t),\nonumber\\ 
&=& \exp[\epsilon(x_t+\imath y_t)]\exp(\imath\theta_t),
\nonumber\\
&=& \exp(\epsilon x_t)\exp[\imath(\theta_t + \epsilon y_t)].
\end{eqnarray}
Notice that $\phi_t\equiv \theta_t + \epsilon y_t$ is a conserved quantity, as
immediately seen from Eq.~(\ref{y}). If we further change the time variable
$t$ to $s$, such that 
\begin{equation}
ds = \exp(\epsilon x_t)dt, 
\label{dsdt}
\end{equation}
we can write
\begin{equation}
\frac{d}{ds} (u_s + \imath v_s) = \exp(\imath \phi_0),
\label{line}
\end{equation}
which is to say $(u_s,v_s)$ is a straight line trajectory of slope
$\tan(\phi_0)$, as announced, and constant speed. This result may indeed
have been anticipated, since the transformation to the new set of variables
coincides with the one for which the iso-kinetic trajectories are
transformed into geodesics of the metric $\exp(2\epsilon x)(dx^2+dy^2)$, see
\cite{DM98}. 

Thus, in terms of the $(u,v,\phi)$ variables,  the time-evolution of
trajectories between collisions is specified according to~: 
\begin{eqnarray}
u_s + \imath v_s &=& u_0 + \imath v_0 + e^{\imath\phi_0} s
\label{utvt}\\
\phi_s &=& \phi_0.
\label{phit}
\end{eqnarray}

The time reparametrization Eq.~(\ref{dsdt}) introduces a new time scale,
which is natural in the new geometry. However it depends on the details of
the trajectory. For future reference, we will make a distinction between
this {\em natural} time $s$ and the {\em physical} time $t$. We integrate
Eq.~(\ref{dsdt}) in order to express the natural time variable, $s$, in
terms of the physical one, $t$~: 
\begin{eqnarray}
s  &=& [(u_0 + 1/\epsilon) \cos\phi_0 + v_0\sin\phi_0][\cosh(\epsilon t) -
  1]\nonumber\\ 
&&+\sqrt{(u_0 + 1/\epsilon)^2 + v_0^2}\sinh(\epsilon t).
\label{st}
\end{eqnarray}

As opposed to the Gaussian iso-kinetic trajectories, the phase-space
volumes are obviously preserved along the trajectories 
Eqs.~(\ref{utvt}-\ref{phit}). The time variables are however different and,
as is clear from Eq.~(\ref{st}), phase-space volumes are not preserved when
the trajectories are parameterized according to the physical time $t$. 

To make the comparison clearer, we first consider the Gaussian iso-kinetic
dynamics, which maps trajectories $(x_0, y_0, \theta_0) \mapsto (x_t, y_t,
\theta_t)$ according to Eqs.~(\ref{x}-\ref{theta}). This map contracts
phase-space volumes, with a rate given by minus the logarithm of the
Jacobian of the application,
\begin{eqnarray}
\Big|\partial_{(x_0,y_0,\theta_0)}(x_t, y_t, \theta_t)\Big|
&=& \partial_{\theta_0} \theta_t,
\nonumber\\
&=& (\cosh\epsilon t + \cos\theta_0\sinh\epsilon t)^{-1},
\nonumber\\
&=& \exp[\epsilon(x_0-x_t)],
\label{pscik}
\end{eqnarray}
{\em i.~e.} $\epsilon(x_t - x_0)$ is the phase-space
contraction rate of the trajectory taking $(x_0,y_0,\theta_0)$ to
$(x_t,y_t,\theta_t)$. 

In the Weyl geometry of the billiard, on the other hand, the phase-space
contraction rate of the physical time trajectory $(u_0, v_0, \phi_0)
\mapsto (u_t, v_t, \phi_t)$ is minus the logarithm of the Jacobian,
\begin{eqnarray}
\lefteqn{\Big|\partial_{(u_0,v_0,\phi_0)}(u_t,v_t,\phi_t)\Big|
= |\partial_{u_0,v_0}(u_t,v_t)|,}&& \nonumber\\
&=& \cosh\epsilon t + \frac{(u_0 + 1/\epsilon) \cos \phi_0 + v_0 \sin
  \phi_0}{\sqrt{(u_0 + 1/\epsilon)^2 + v_0^2}}  \sinh \epsilon t, 
\nonumber\\
&=& \frac{\sqrt{(u_t+1/\epsilon)^2+v_t^2}}
{\sqrt{(u_0+1/\epsilon)^2+v_0^2}}.
\label{pscconf}
\end{eqnarray}
This is the inverse of Eq.~(\ref{pscik}), as easily checked. Therefore, so
long as the time scales are the same, the phase-space volume contraction
changes sign, going from the dynamics of the iso-kinetic billiard to that of
the W-flow. Moreover minus the logarithm of the latter expression yields
$\epsilon(u_0 - u_t)$ only to first order in $\epsilon$. The reason for
this is connected to the transformation law of $x$,
Eq.~(\ref{confmapexp})~: $x = 1/\epsilon \log\sqrt{(u+1/\epsilon)^2 +
  v^2}$. 

As remarked by Dettmann and Morriss \cite{DM98}, the periodic boundary
conditions break the Hamiltonian structure of the billiard in the Weyl
geometry. Indeed, they do not preserve phase-space volumes as we now
show.  

In order to analyze the billiard trajectories in the Weyl geometry, we must
first consider the transformation of the unit cell under the conformal map,
Eq.~(\ref{confmapexp}), as shown on the right panel of
Fig.~\ref{fig.gik}. It is straightforward to check the rectangular cell is 
mapped to a trapezoidal figure with curved lateral sides, and the disks to
slightly distorted, flattened ones. Here are the parametric equations of
these elements~: 
\begin{eqnarray*}
\hspace{-1cm}\mathrm{upper/lower\ sides:}
&\ & \Big\{
u + \imath v 
= \epsilon^{-1}[\exp(\epsilon t \pm \imath\sqrt{3}\epsilon/2) - 1], 
\: -1/2\leq t\leq 1/2
\Big\},\\
\hspace{-1cm}\mathrm{right/left\ sides:}
&\ & \Big\{
u + \imath v =  
\epsilon^{-1}
[\exp(\pm \epsilon/2 + \imath \sqrt{3}\epsilon t) - 1], \: -1/2\leq t\leq
1/2 \Big\},\\
\hspace{-1cm}\mathrm{central\ disk:}
&\ &
\Big\{
u + \imath v = \epsilon^{-1}
\{\exp[\epsilon \sigma \exp (\imath \phi)] - 1\},\:
0\leq\phi\leq2\pi\Big\},\\
\hspace{-1cm}\mathrm{outer\ disks:}
&\ &
\Big\{
u + \imath v = \epsilon^{-1}
\{\exp\big[\epsilon\big(\pm 1/2 \pm \imath \sqrt{3}/2 +
\sigma \exp (\imath \phi)\big)\big] - 1\},\\
&&\hspace{7cm} 0\leq\phi\leq2\pi\Big\}.
\end{eqnarray*}


Notice that the upper and lower borders are oblique lines of slopes $\pm
\sqrt{3}\epsilon/2$, while the left and right borders are concentric
arc-circles of center $(-1/\epsilon, 0)$ and respective radii
$\exp(\pm\epsilon/2)/\epsilon$. Thus, the periodic 
boundary conditions induce phase-space expansion/contraction as the variables
$u$ and $v$ are rescaled by a factor $\exp(\pm\epsilon)$ when the
trajectory crosses from left to right or right to left. 

In order to analyze the ergodic properties of the W-flow, we turn to the
definition of the Birkhoff coordinates of the Weyl billiard in the periodic
cell, which specify the evolution of trajectories from one collision to the
next (including collisions with the boundaries). Three coordinates are
necessary here~: $\varsigma$, which denotes the arc-length along the
boundaries of the unit cell (walls and disks), $\omega$, the modulus of the
velocity vector, and $\varpi$, by which we denote the sinus of the angle
between the outgoing trajectory and the normal to the obstacles' boundary,
that is the tangent component of the  unit vector in the direction of the
velocity. As a result, in the stationary state, so long as the regime is
hyperbolic, there are three Lyapunov exponents which characterize the
chaoticity of the Birkhoff map. We denote them by $\lambda_1 > \lambda_2 >
\lambda_3$, where $\lambda_1$ and $\lambda_3$ are associated to the
$(\varsigma, \varpi)$-dynamics, and $\lambda_2$ to the $\omega$-dynamics,
corresponding to the direction of the flow, and which is zero as it will
turn out.

We first consider the dynamics of $\omega$. Here we have to make a distinction
between the two time parameterizations. If, on the one hand, we integrate trajectories with
respect to the physical time $t$, the modulus of the velocity is typically
expanded in the bulk according to Eq.~(\ref{pscconf}), while it is
contracted at the opposite rate by the periodic boundary conditions. That
the particle's velocity must be rescaled by a factor $e^{\pm\epsilon}$ at
the periodic boundaries can be seen from the expression of the
velocity, given by Eq.~(\ref{dsdt}), here expressed in terms of the
variables $u, v, \phi$,
\begin{eqnarray}
\omega_t = \frac{d}{dt}s &=& \epsilon \Big\{
[(u_0 + 1/\epsilon) \cos \phi_0 + v_0 \sin \phi_0] \sinh(\epsilon t)
\nonumber\\
&& \phantom{\epsilon}
+ \sqrt{(u_0 + 1/\epsilon)^2 + v_0^2} \cosh(\epsilon t)\Big\}.
\label{v}
\end{eqnarray}
This rescaling yields a contribution to $\lambda_2$ equal to\footnote{For
  definiteness we assume the particle density to be normalized to unity. }
$- \epsilon J_c$, where $J_c$ is the (signed) average number of boundary
crossings per
collision, {\em i.~e.} the average drift velocity per collision, a positive
quantity. The other contribution occurs from the bulk, because
of the dissipation that takes place along the trajectories, and is given by
the logarithm of Eq.~(\ref{pscconf}). As we noted by comparing it to
Eq.~(\ref{pscik}), this contribution is opposite to the phase-space
contraction rate of the iso-kinetic billiard along the image trajectory.
The latter quantity is equal to the entropy production rate of the
iso-kinetic Lorentz gas, given by the product of the applied
force by the resulting drift current \cite{CELS93}, {\em i.~e.} $\epsilon
J_c$. Thus the bulk contribution to $\lambda_2$ is identical but opposite
to the contribution due to the boundary conditions. Therefore, 
\begin{equation}
\lambda_2 = 0.
\label{lambda2}
\end{equation}
If, on the other hand, we integrate the trajectories with respect to the
natural time, {\em i.~e.} with respect to $s$, the dynamics
of $\omega$ are trivial. Indeed the velocity
is constant in the bulk and remains so under the periodic boundary
conditions. This is because the $s$ increments are rescaled at the
boundaries identically to the length increments, as can be seen from
Eq.~(\ref{v}). Thus, trivially, we again retrieve $\lambda_2 = 0$.

As noticed earlier, Eqs.~(\ref{utvt}-\ref{phit}), the dynamics of
$\varsigma$ and $\varpi$ preserves the phase-space volumes in the 
bulk. This remains so whether the time variable is $s$ or $t$. Since the
trajectories follow straight lines, the evolution of these two variables
from one collision event to the next is the same, whether the 
velocity changes in the bulk or remains constant. However, as noted
earlier, phase-space volumes are not preserved when the periodic boundary
conditions apply. The arc-length coordinate is contracted by a factor
$e^{\pm\epsilon}$. We can therefore write minus the sum of the two
Lyapunov exponents associated to the dynamics of $\varsigma$ and $\varpi$
as 
\begin{equation}
-\lambda_1 - \lambda_3 = \epsilon J_c,
\label{lambda13}
\end{equation}

The total phase-space contraction rate is the sum of Eqs.~(\ref{lambda2})
and (\ref{lambda13}), equal to the product of the current multiplied by the
amplitude of the external field. This is obviously the same result as
obtained for the iso-kinetic Lorentz billiard \cite{CELS93}, which we did
expect since the winding number, and hence the current $J_c$, are invariant
under the conformal map. Table \ref{tbl.psc} summarizes our
findings. Notice that the current $J_c$ is here measured with respect to
the collision dynamics. The identification between the phase-space
contraction rate and the entropy 
production per unit time (in the regime of linear transport) can be done by
substituting the average current per collision, $J_c$, by the corresponding
average current per unit time, which we denote by $J_t$. This is
straightforward since the two quantities differ only by a conversion
factor, the average free flight time, or average time between
collisions. On the other hand, it should be mentioned the conversion to
a current measured in the units of $s$ is not meaningful since this
integration variable changes from one trajectory to another. Each
trajectory travels with its own clock, which impedes a statistical
interpretation. 

\begin{table}
\begin{center}
\begin{tabular}{||l||c|c|c||}
\hline
Dissipation&Bulk&Boundaries&Total\\
\hline
GIK LG& $\epsilon J_c$&$0$&$\epsilon J_c$\\
W-flow ($t$)& $-\epsilon J_c$&$2\epsilon J_c$&$\epsilon J_c$\\
W-flow ($s$)&$0$&$\epsilon J_c$&$\epsilon J_c$\\
\hline
\end{tabular}
\end{center}
\caption{Comparison of the different contributions to the phase-space
  contraction rate for the Gaussian iso-kinetic Lorentz gas and its two
  equivalent representations in the Weyl geometry, whether the integration
  variable is $t$, the physical time, or $s$, the natural time,
  Eq.~(\ref{st}). In the former case, the contraction of the velocity at
  the periodic boundaries compensates the expansion of phase-space volumes
  in the bulk.} 
\label{tbl.psc}
\end{table}

To conclude, we have shown how the transformation of the Gaussian
iso-kinetic Lorentz channel to a Weyl geometry maps the trajectories into
straight lines with constant speed, so long as we rescale the time variable
according to Eq.~(\ref{dsdt}). The geometry of the cell is modified under
this transformation. The vertical walls of the Lorentz channel are here
replaced by concentric circles whose radii differ by a factor given by the
exponential of the external field. The disks are also deformed and can even
have concavities when the field strength becomes too large ($\epsilon >
1/\sigma$, which corresponds to the transition to non-hyperbolic regime,
as shown in \cite{W00}).  

The average phase-space contraction rate, equal (at least in the small
$\epsilon$ regime) to the entropy production rate, is here due to the
application of the periodic boundary conditions. A remarkable consequence
of this, is that the calculation of the phase-space contraction, as well as
its identification to the entropy production rate, are immediate.

It should be pointed out that the expression of the phase-space contraction
rate, Eq.~(\ref{lambda13}), makes no reference to the actual time scale
since it relies solely on the collision dynamics. It is in particular
transparent to the change of time scales, Eq.~(\ref{dsdt}). However, as we
argued, only the physical time is relevant to statistical averages. In this
system the velocities change with the physical time and get rescaled at the
periodic boundaries. The velocity dynamics of the natural time are on the
other hand trivial. Thus one might wonder what happens when $s$ is taken as
the physical time, in which case we lose the connection with the Gaussian
iso-kinetic dynamics. This is what we do in \cite{BG}, where we 
discuss an expanding billiard model similar to this one, but for which the
physical time dynamics is trivial between collisions and preserves
phase-space volumes in the bulk. In this case, the speed variable must be
rescaled at the periodic boundaries, and is associated to a third Lyapunov
exponent, here negative. The mobility changes by a factor of two,  
and one nevertheless retrieves an identification between entropy production
and phase-space contraction rates in the linear transport regime.

\ack
The authors wish to thank N. I. Chernov and C. Liverani for insightful
discussions. FB acknowledges financial support from Fondecyt
project 1060820 and FONDAP 11980002. TG  is financially supported by the
Fonds National de la Recherche Scientifique. This collaboration was
partially supported through grant ACT 15 (Anillo en Ciencia y Tecnologia).

\end{document}